\documentclass[conference]{IEEEtran}
\IEEEoverridecommandlockouts
\usepackage{cite}
\usepackage{amsmath,amssymb,amsfonts}
\usepackage{algorithmic}
\usepackage{graphicx}
\usepackage{textcomp}
\usepackage{xcolor}
\usepackage{subfigure}
\usepackage{url}
\usepackage{multirow}
\usepackage{soul}
\usepackage{tabularx,booktabs}

\def\BibTeX{{\rm B\kern-.05em{\sc i\kern-.025em b}\kern-.08em
		T\kern-.1667em\lower.7ex\hbox{E}\kern-.125emX}}
		
\begin{document}
	
	\title{On the Performance of the Spatial Reuse Operation in IEEE 802.11ax WLANs
		\thanks{This  work  has  been  partially  supported  by  the Spanish Ministry of Economy and Competitiveness under the Maria de Maeztu  Units  of  Excellence  Programme (MDM-2015-0502), by WINDMAL PGC2018-099959-B-I00 (MCIU/AEI/FEDER,UE), by the Catalan Government under SGR grant for research support (2017-SGR-11888),  and  by a Gift from the Cisco University Research Program (CG\#890107, Towards Deterministic Channel Access in High-Density WLANs) Fund, a corporate advised fund of Silicon Valley Community Foundation. The work by S. Barrachina-Mu\~noz is supported by an FI grant from the Generalitat de Catalunya.}}
	
	\author{\IEEEauthorblockN{1\textsuperscript{st} Francesc Wilhelmi}
		\IEEEauthorblockA{\textit{Wireless Networking Group} \\
			\textit{Universitat Pompeu Fabra}\\
			Barcelona, Spain  \\
			francisco.wilhelmi@upf.edu}
		\and
		\IEEEauthorblockN{2\textsuperscript{nd} Sergio Barrachina-Mu\~noz}
		\IEEEauthorblockA{\textit{Wireless Networking Group} \\
			\textit{Universitat Pompeu Fabra}\\
			Barcelona, Spain  \\
			sergio.barrachina@upf.edu}
		\and
		\IEEEauthorblockN{3\textsuperscript{rd} Boris Bellalta}
		\IEEEauthorblockA{\textit{Wireless Networking Group} \\
			\textit{Universitat Pompeu Fabra}\\
			Barcelona, Spain \\
			boris.bellalta@upf.edu}
	}
	
	\maketitle
	
	\begin{abstract}
	The Spatial Reuse (SR) operation included in the IEEE 802.11ax-2020 (11ax) amendment aims at increasing the number of parallel transmissions in an Overlapping Basic Service Set (OBSS). However, many unknowns exist about the performance gains that can be achieved through SR. In this paper, we provide a brief introduction to the SR operation described in the IEEE 802.11ax (draft D4.0). Then, a simulation-based implementation is provided in order to explore the performance gains of the SR operation. Our results show the potential of using SR in different scenarios covering multiple network densities and traffic loads. In particular, we observe significant improvements on the channel utilization when applying SR with respect to the default configuration, thus allowing to increase the throughput and reduce the delay. Interestingly, the highest improvements provided by the SR operation are observed in the most pessimistic situations in terms of network density and traffic load.
	\end{abstract}
	
	
	\section{Introduction}
	
	The IEEE 802.11ax (11ax) amendment, which official publication is due to be released in June 2020, is expected to lay the groundwork of Next-Generation (NG) Wireless Local Area Networks (WLANs). One of the main goals of this amendment is to improve network efficiency by increasing the number of parallel transmissions in an Overlapping Basic Service Set (OBSS). To that purpose, the Spatial Reuse (SR) operation is introduced along with other techniques to boost the performance of NG WLANs, from which we highlight Orthogonal Frequency-Division Multiple Access (OFDMA) or Downlink/Uplink Multi-User  Multiple-Input-Multiple-Output (DL/UL MU-MIMO) \cite{bellalta2016ieee}.
	
	The SR operation is based on sensitivity adjustment together with Transmission Power Control (TPC). In particular, a specific OBSS Packet Detect (OBSS/PD) threshold is employed for the detected OBSS transmissions (also referred to as inter-BSS transmissions), so that channel utilization can be enhanced. Moreover, in order not to affect any ongoing transmission, a node applying SR must limit its transmit power as a function of the OBSS/PD threshold.
	
\begin{figure}[ht!]
		\centering
		\includegraphics[width=0.6\columnwidth]{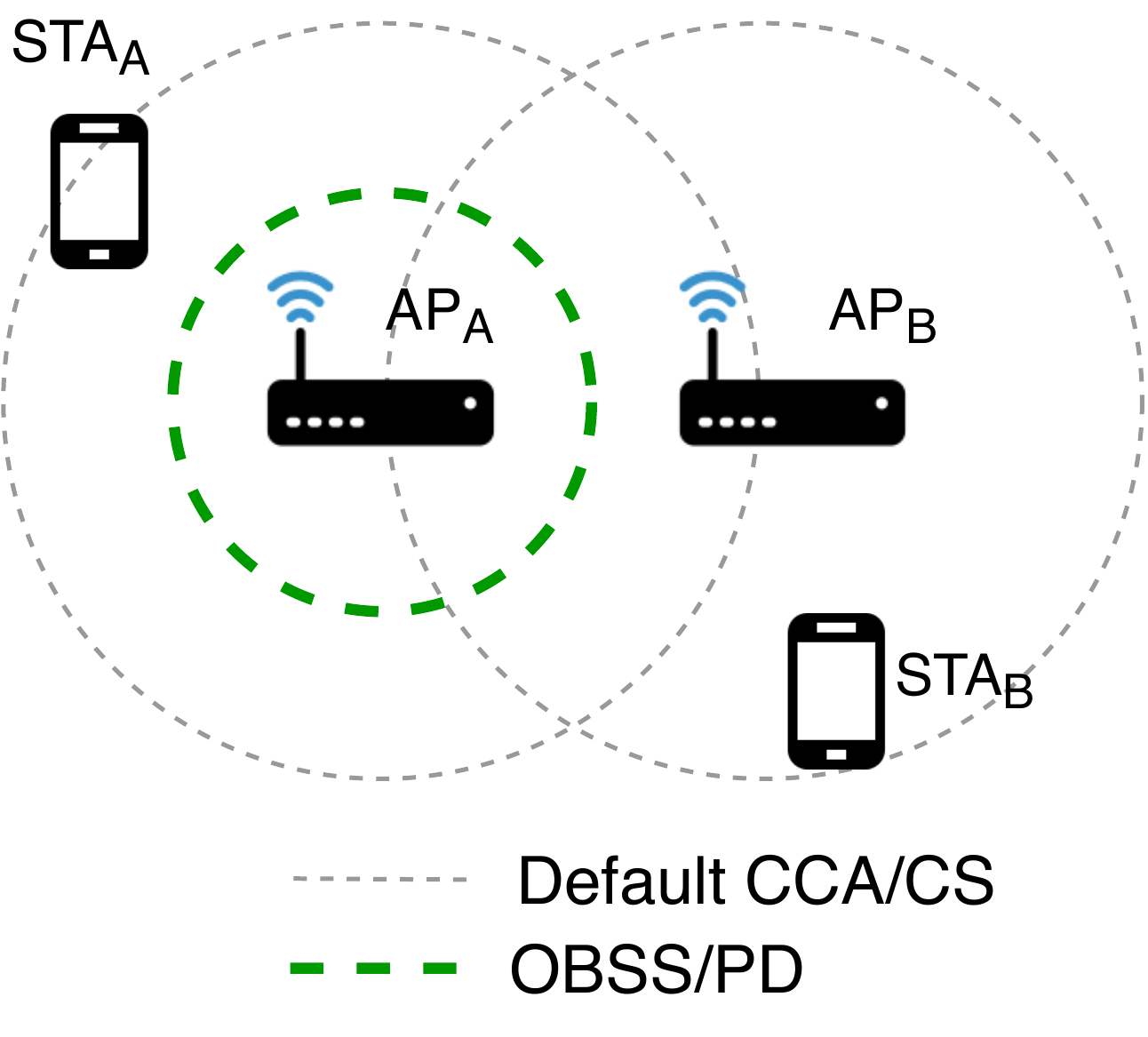}
		\caption{Example of the potential of the 11ax SR operation.}
		\label{fig:example_sr}
	\end{figure}

	Fig.~\ref{fig:example_sr} depicts a use case where the SR operation could potentially improve the network efficiency of an OBSS. Notice that the dashed lines in the figure indicate the carrier sense area of each device, provided that the transmit power of the others is fixed and that the same channel is used. As illustrated, the default Clear Channel Assessment Carrier Sense (CCA/CS) threshold would not allow simultaneous transmissions to be held between Access Points \textit{A} and \textit{B} (AP$_A$ and AP$_B$). In that case, each device should defer its transmission when the other occupies the channel, due to the application of the Carrier Sense Multiple Access with Collision Avoidance (CSMA/CA) protocol. Nevertheless, by properly increasing the OBSS/PD threshold of any AP (e.g., as illustrated for AP$_A$), both devices would be able to transmit at the same time, thus improving the utilization of the channel.
	
	Despite SR is expected to bring significant performance gains to WLANs, its actual benefits are still unknown. First, the new kind of inter-WLAN interactions that the operation generates is unexplored, as well as their impact on the network. Moreover, the improvement achieved by applying SR on the performance (e.g., throughput, delay) is hindered by the trade-off between the number of parallel transmissions and their duration. Note that increasing the OBSS/PD threshold (i.e., using a more aggressive configuration) entails decreasing the transmission power, which may result in using a lower Modulation and Coding Scheme (MCS), or even experiencing an increased packet error rate. The effects of increasing the OBSS/PD threshold, and hence decreasing the transmission power, are summarized in Table \ref{tbl:effects_sr}. As shown, an increase in the OBSS/PD threshold entails a higher probability of accessing to the channel since the number of sensed inter-BSS transmissions can be potentially reduced (this is equivalent to reducing the exposed-node probability). In contrast, the number of hidden nodes can potentially increase as the carrier sense area is being reduced.
	
    \begin{table}[]
    	\caption{Effect of increasing the OBSS/PD threshold and the transmission power.}
    	\label{tbl:effects_sr}
    	\resizebox{\columnwidth}{!}{\begin{tabular}{|c|c|c|c|c|}
    		\hline
    		& \begin{tabular}[c]{@{}c@{}}Data \\ rate\end{tabular} 
    		& \begin{tabular}[c]{@{}c@{}}Channel access\\ probability\end{tabular} &  \begin{tabular}[c]{@{}c@{}}Hidden-node\\ probability\end{tabular} & \begin{tabular}[c]{@{}c@{}}Exposed-node\\ probability\end{tabular} \\ \hline
    		\begin{tabular}[c]{@{}c@{}}OBSS/PD $\uparrow$\\ (Tx Power $\downarrow$)\end{tabular} & $\downarrow$ & $\uparrow$ & $\uparrow$ & $\downarrow$ \\ \hline
    	\end{tabular}}
    \end{table}
	
    In this work, we shed light on the performance of the 11ax SR operation and highlight the situations in which it is worth using it. The main contributions of this paper are as follows:
	\begin{itemize}
		\item We provide a summary of the OBSS PD-based SR operation included in draft version D4.0 of the IEEE 802.11ax amendment, which is, to the date of publishing this article, under the initial sponsor ballot phase.
		\item We present an implementation of the aforementioned operation in the 11ax-based Komondor simulator \cite{komondor}.\footnote{All the source code of Komondor is open and free to use (Github repository: \url{https://github.com/wn-upf/Komondor}), with the aim of encouraging potential collaborations with any interested researcher.}
		\item We evaluate the performance of the SR operation through simulations, and assess its potential for next-generation wireless networks. Different network densities and traffic loads are considered for covering the analysis of multiple use cases.
	\end{itemize}
	
	\section{Related Work}
	\label{section:related_work}
	
	The SR operation has been previously surveyed and evaluated in \cite{bellalta2016ieee,khorov2018tutorial,sr_evaluation_1,sr_evaluation_3,sr_evaluation_2}. However, these works refer to previous draft versions of the amendment (D1.0 o D2.0), which has undergone significant modifications in its current version (D4.0).

	First, the Task Group ax (TGax) presented some preliminary results for cellular-type scenarios in \cite{sr_evaluation_1}. In particular, significant gains were shown when combining BSS Coloring and Dynamic Sensitivity Control (DSC) \cite{smith2017dynamic}. A further analysis was then provided in \cite{sr_evaluation_3} for office scenarios, which also showed that gains were only achieved in dense deployments. Nonetheless, the simulations conducted in that work were obtained from a customized system and link level integrated simulation platform, from which no validation was provided.
	
	The authors in \cite{sr_evaluation_2} provided a thorough performance evaluation of the SR operation, in addition to several other features included in the 11ax amendment. To that purpose, they proposed their own simulation platform for IEEE 802.11ax called SLISP, which mostly focuses on the MAC of the 11ax. Based on that, the SR operation was evaluated in indoor and outdoor scenarios containing multiple BSS. While important gains were shown in indoor deployments (especially for downlink traffic), a moderate gain was observed in outdoor situations.
	
	As shown, few works attempt to provide a performance evaluation of the SR operation through simulations. The main cause lies in the novelty of the mechanism. Accordingly, there is a lack of reliable simulation platforms that include 11ax features. To the date of publishing this article, SR is still under development for ns-3.\footnote{All the new developments in ns-3 are published in the following repository: \url{https://gitlab.com/nsnam/ns-3-dev}} Due to the lack of simulation tools including the 11ax SR operation, in this work we provide an implementation of the SR operation in the Komondor simulator.\footnote{The validation of the Komondor simulator against ns-3 can be found in~\cite{komondor}.} Moreover, our results are gathered based on the newest draft version (D4.0).
	
	\section{Spatial Reuse Operation}
	\label{section:sr_operation}
	
	\begin{figure*}[ht!!!!]
		\centering
		\subfigure[Scenario]{\label{fig:fig_2}\includegraphics[width=0.6\columnwidth]{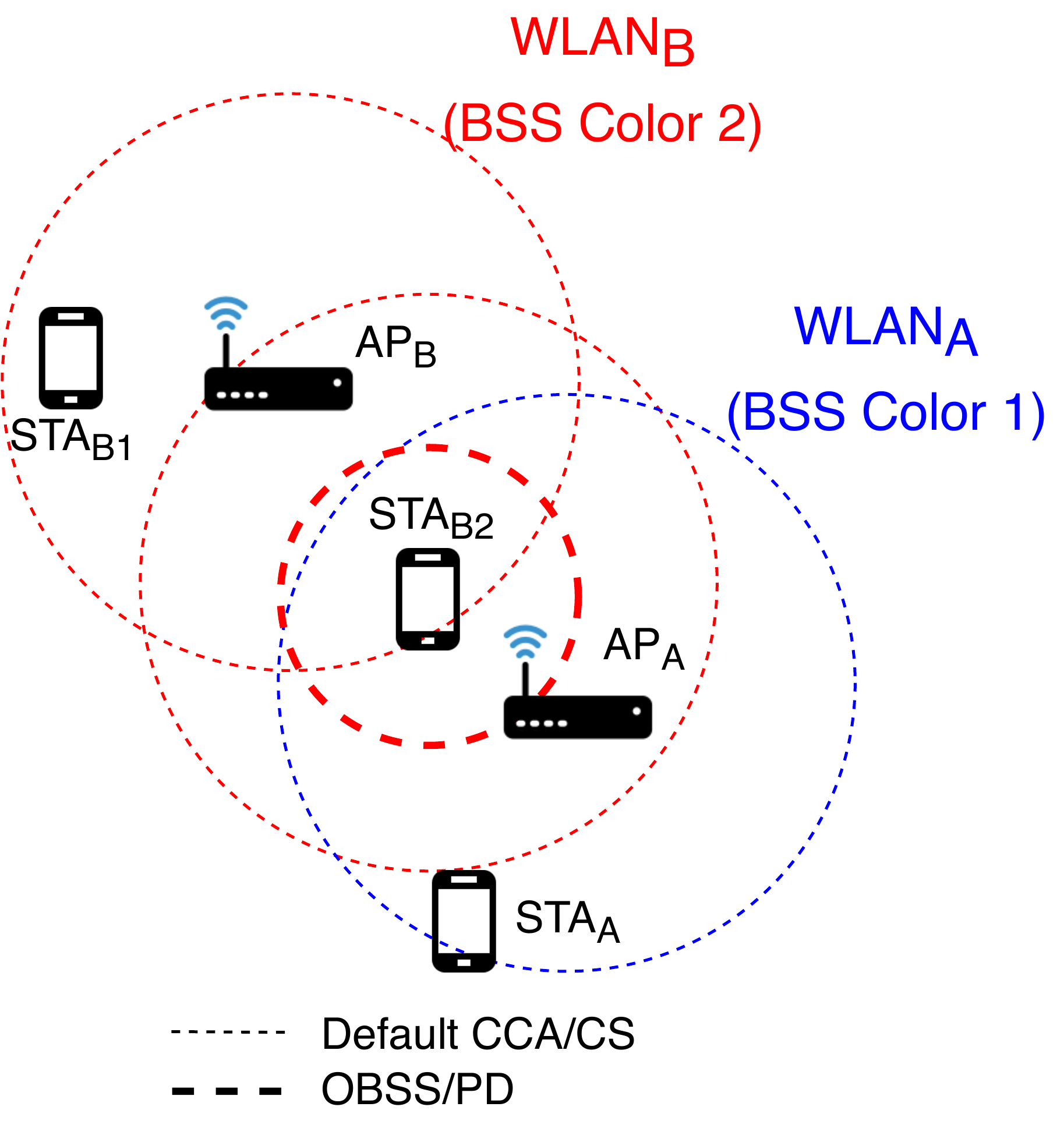}}
		\subfigure[Packets exchange]{\label{fig:fig_2b}\includegraphics[width=0.7\columnwidth]{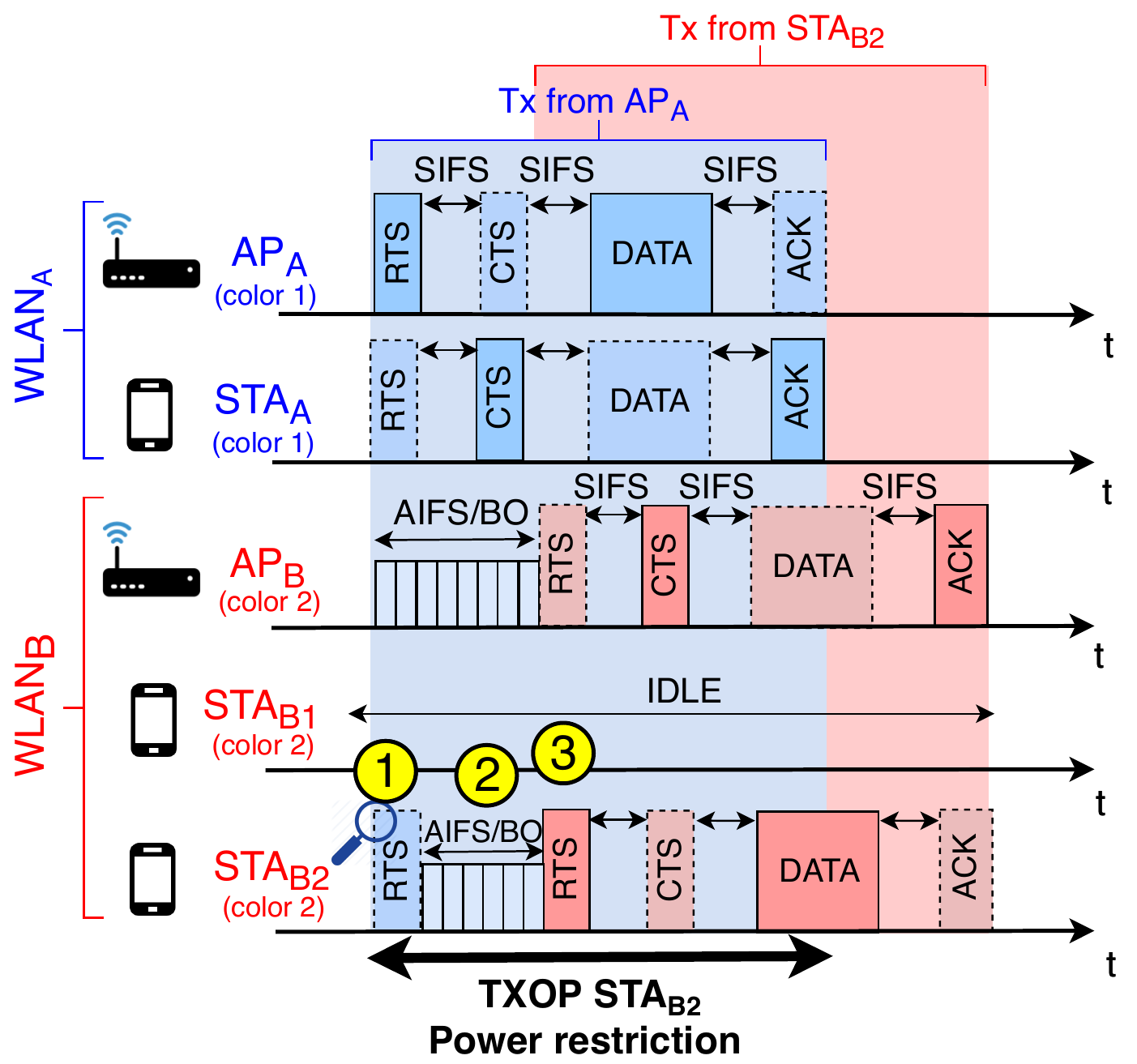}}
		\caption{Example of applying the OBSS PD-based SR operation.}
		\label{fig:scenario_example_2}
	\end{figure*}
	
	The 11ax SR operation is divided into two different and independent mechanisms. On the one hand, we find the OBSS PD-based SR operation, whereby 11ax devices - a.k.a High Efficiency (HE) nodes\footnote{By 11ax node, we may refer indistinctly to an HE STA or an HE AP.} - can detect SR opportunities from inter-BSS transmissions by using a more aggressive CCA policy. On the other hand, the Spatial Reuse Parameter (SRP)-based SR operation performs similarly but only taking advantage of trigger-based communications~\cite{bellalta2019ap}. Throughout this document, we will exclusively refer to the first mechanism (i.e., OBSS PD-based SR) because of the development cost of building trigger-based transmissions required for the SRP-based SR operation. In addition, the slow adoption of 11ax in WLANs would prevent using full scheduling transmissions schemes, in favor of CSMA/CA ones. Notwithstanding, both mechanisms are expected to lead to similar results since the procedure of adjusting the OBSS/PD is similar.
	
	\subsection{BSS Coloring and Spatial Reuse Groups}
	
	The whole SR operation is based on identifying the source of a given transmission, i.e., inter-BSS frame detection. The idea is that HE nodes can rapidly decode the MAC headers of a certain transmission and determine its origin. Then, a more aggressive OBSS/PD threshold can be employed to increase the probability of accessing to the channel.
	
	For the fast packet source identification, two concepts are introduced, which stand for BSS Coloring and Spatial Reuse Groups (SRG). On the one hand, the BSS Color field is included in the MAC headers\footnote{The BSS Color is carried in the HE-SIG-A field, which is present in every Physical Layer Convergence Procedure (PLCP) Protocol Data Unit (PPDU).} to uniquely identify different WLANs belonging to an OBSS. In case of detecting a color collision, the affected WLANs must change their BSS Color. On the other hand, SRGs can be formed by a set of overlapping WLANs. The SRG field is present in control frames such as Beacons, Probe responses, or (Re)Association responses.\footnote{Unlike the BSS Color, the SRG is included in the Spatial Reuse Parameter Set (SRPS) element. A bitmap is stored by each 11ax node applying SR, which maps the set of BSS Colors that belong to a certain SRG.} In this case, a specific OBSS/PD threshold can be used for transmissions within the same SRG.
		
	\subsection{General Constraints}
	The 11ax amendment includes a set of constraints on defining the OBSS/PD threshold to be used for detecting SR opportunities. In particular, the OBSS/PD value cannot exceed the following upper bound:
	\begin{align}\nonumber \text{OBSS/PD} \leq & \max\Big(\text{OBSS/PD}_{\min}, \min\big(\text{OBSS/PD}_{\max},\\ & \text{OBSS/PD}_{\min} + (\text{TX\_PWR}_{\text{ref}}-\text{TX\_PWR})\big)\Big), \nonumber \end{align}
	where $\text{OBSS/PD}_{\min}$ and $\text{OBSS/PD}_{\max}$ are set to $-82$ dBm and $-62$ dBm, respectively, the reference power $\text{TX\_PWR}_{\text{ref}}$ is set to 21 or 25 dBm, according to the
	capabilities of the device,\footnote{The $\text{TX\_PWR}_{\text{ref}}$ can be set to either 21 or 25 dBm, depending on the transmission capabilities of the HE node with regards to the highest supported number of spatial streams (NSS).} and $\text{TX\_PWR}$ is the transmission power in dBm.
	
	In order to regulate the transmissions held during SR-based opportunities, the transmission power is limited according to the OBSS/PD threshold used for detecting those opportunities. In case that $\text{OBSS/PD} \leq \text{OBSS/PD}_{\min}$, the transmission power is unconstrained. Otherwise, the maximum allowed transmission power $\text{TX\_PWR}_{\max}$ is given by:
	\begin{equation}
	\resizebox{.9\columnwidth}{!}{$\text{TX\_PWR}_{\max} = \text{TX\_PWR}_{\text{ref}} - (\text{OBSS/PD} -\text{OBSS/PD}_{\min})$}
	\label{eq:power_restriction}
	\end{equation}
		
	\subsection{Example of the OBSS PD-based Spatial Reuse Operation}

	In order to illustrate the concepts described above, let us consider the scenario depicted in Fig.~\ref{fig:fig_2}. As shown, there is a device, namely STA$_{B2}$, which, by using the default configuration, is prone to suffer from flow starvation as a result of the OBSS interference. Nevertheless, the OBSS PD-based SR operation allows STA to overcome the aforementioned interference, thus gaining access to the channel. This is illustrated in Fig.~\ref{fig:fig_2b}, where inter-BSS transmissions are ignored by STA$_{B2}$ when using the SR operation. In marker~1 (shown in yellow), STA$_{B2}$ inspects the Request to Send (RTS) frame sent by AP$_A$, which is identified as an inter-BSS transmission. Accordingly, it uses a more aggressive OBSS/PD threshold, which allows the backoff procedure to be resumed (marker 2). Finally, STA$_{B2}$ starts its own transmission by taking advantage of the detected SR-based opportunity (marker 3). However, a power restriction is applied, thus decreasing the MCS and increasing the data transmission time, as a consequence.
	
	\section{Implementation of OBSS PD-based SR in Komondor}
	\label{section:komondor}
	
	The Komondor simulator was conceived, among other purposes, to allow the low-complexity integration of novel mechanisms included in new IEEE 802.11 standards, which are not yet available or validated in other well-known simulators such as ns-3. Therefore, Komondor can serve as a first step towards analyzing novel features that will potentially shape future wireless networks. In this Section, we briefly introduce the implementation conceived for the SR operation.\footnote{The code used for the simulations of this paper can be found in pre-release v3.0 (\url{https://github.com/wn-upf/Komondor/releases/tag/3.0}).}
	
\begin{figure}[h!!!!]
		\centering
		\includegraphics[width=0.95\columnwidth]{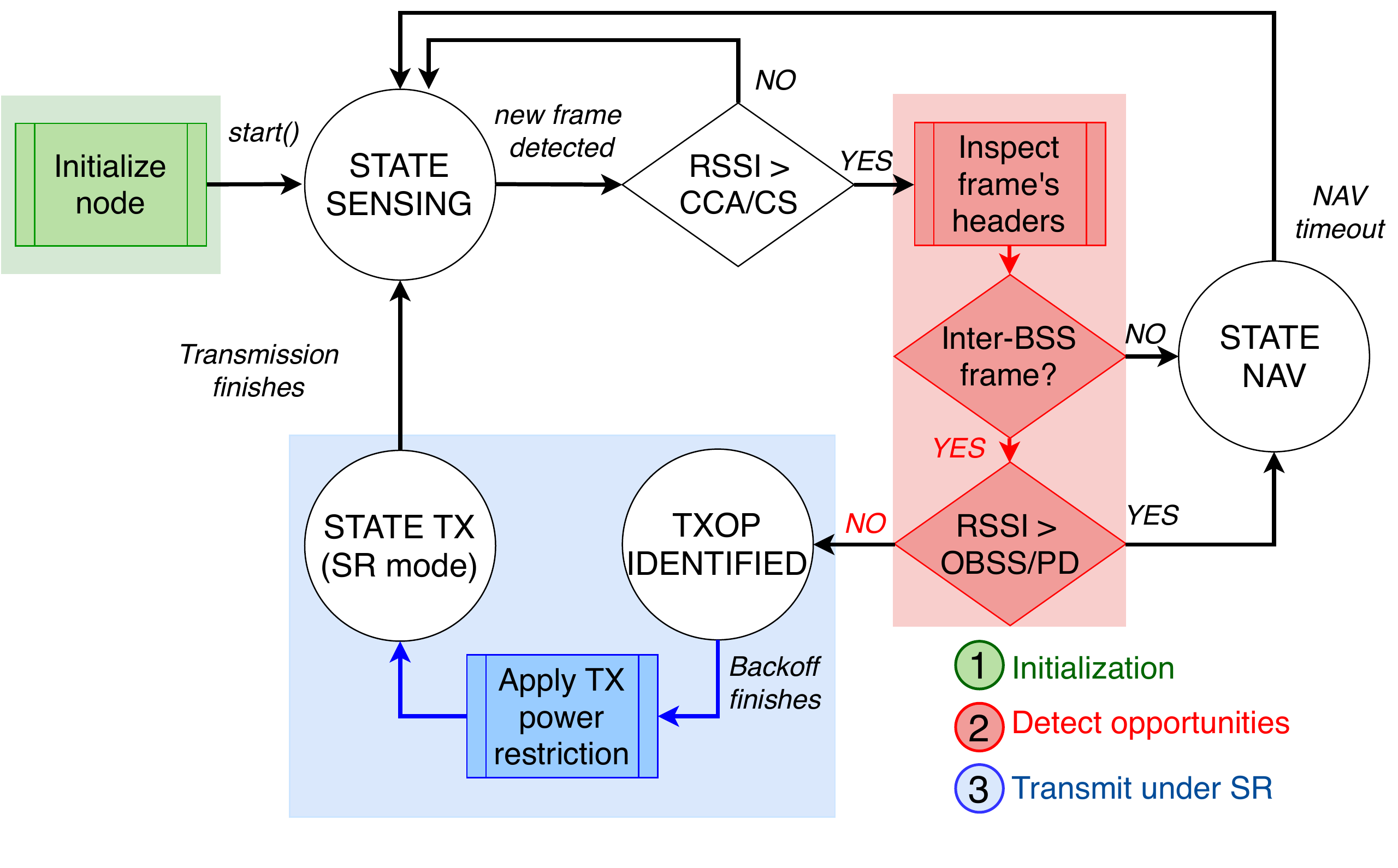}
		\caption{Flowchart of the SR implementation for a given HE node when detecting a single inter-BSS transmission.}
		\label{fig:implementation_overview}
	\end{figure}

	Fig.~\ref{fig:implementation_overview} shows a flowchart that summarizes the SR implementation for a given HE node in case of detecting a single inter-BSS transmission. The most important groups of functionalities (highlighted with numbers in the figure) are described in detail in the following subsections.
	
	\subsection{Initialization}
	First of all, any node that applies SR must indicate support for it. In addition, the parameters related to the SR operation should be set:
	\begin{itemize}
		\item \textbf{BSS Color}: identifier of the BSS to which the node belongs to. The BSS Color identifies WLANs uniquely.
		\item \textbf{Spatial Reuse Group (SRG)}: identifier of the group to which the node belongs.
		\item \textbf{Non-SRG OBSS/PD}: the sensitivity threshold to be used for generic inter-BSS transmissions.
		\item \textbf{SRG OBSS/PD threshold}: the sensitivity threshold to be used for inter-BSS transmissions that are originated by nodes belonging to the same SRG.
	\end{itemize}
	
	The Komondor simulator simplifies the PHY layer for the sake of efficiency, so that particular focus is put on the MAC. Moreover, operations related to management and control are simplified too. According to that, the initialization of nodes is logically performed at the beginning of the simulation, instead of simulating the actual exchange of control frames between nodes. In particular, the AP of a given WLAN is responsible for notifying the initial SR configuration and any potential change to its associated STAs.

	\subsection{Detection of SR-based Opportunities}
	
	Once the simulation starts and nodes begin to exchange packets, it is possible to detect SR-based opportunities from inter-BSS frames. For that, a certain HE node must first analyze the headers of any detected frame and rapidly identify its source. During this stage, the HE node will assess whether the transmitter belongs to the same WLAN (intra-BSS) or not (inter-BSS). Moreover, in case of being of kind inter-BSS, the frame is sub-categorized into SRG or non-SRG, according to the groups established during initialization.
	
	In case of detecting an intra-BSS transmission, the default CCA/CS threshold is used. Otherwise, the corresponding OBSS/PD threshold (non-SRG or SRG) is applied. In accordance with that, the power received P$_\text{rx}$ from the incoming transmission is used to identify potential SR-based opportunities. In particular, the following two conditions must hold to identify an opportunity: 1)~P$_\text{rx}$ $\geq$ CCA/CS, to guarantee the correct decoding of the MAC header, and 2) P$_\text{rx}$ $<$ OBSS/PD, to trigger the opportunity.
	
	\subsection{Transmit under the SR mode}
	When detecting an SR-based opportunity, an HE node detects the channel as idle, which allows decreasing the backoff. Once the backoff counter is over and the node is about to transmit, a transmit power limitation is applied \eqref{eq:power_restriction}. Finally, once the HE node finishes its SR-based transmission, it returns to the default sensing state, where channel access is scheduled according to the legacy CCA/CS threshold.
	
	It is important to notice that several SR-based opportunities can be detected before transmitting, due to the multiple receptions of different inter-BSS frames. In that case, the most restrictive power limitation must be applied. 
	
	\section{Simulation Setup}
	\label{section:simulation_setup}
	In this Section, we depict the simulation setup that has been considered for evaluating the performance of the SR operation.
	
	\subsection{Channel Model}
	Path-loss effects are characterized according to the TMB 5GHz indoor model for IEEE 802.11ac/11ax WLANs \cite{tmb}. In particular, the path-loss $\text{PL}_\text{TMB}$ between a transmitter $i$ and a receiver $j$ that are separated by $d_{i,j}$ meters is given by:
	\begin{equation}
	\text{PL}_\text{TMB}(d_{i,j}) = L_0 + 10 \cdot \gamma \cdot \log_{10}(d_{i,j}) + k \cdot \overline{W} \cdot d_{i,j},
	\label{eq:tmb}
	\nonumber
	\end{equation} 
	where  $L_0$ is the path-loss
    intercept, $\gamma$ is the path-loss exponent, $k$ is the attenuation factor that characterizes obstacles, and $\overline{W}$ is the average number of wall obstacles per meter.
	
	\subsection{Traffic Generation and Data Rate}
	Only downlink transmissions are considered for the sake of capturing inter-AP interactions. Hence, a UDP traffic generator is attached to every AP. All traffic generators randomly produce packets at the same average traffic load $\ell$, which varies depending on the scenario. The packets arrival process to the APs is modeled through the well-known Poisson distribution.
	
	The rate at which data is transmitted is based on the MCS modes defined in the 11ax amendment, which are selected according to the link quality between the transmitter (the AP) and the receiver (the STA). The highest achievable data rate (135 Mbps) is achieved when using modulation 1024-QAM at a coding rate of 5/6.
	
	\subsection{Throughput Calculation and Reception Model}
	
	Nodes operate under the CSMA/CA protocol, and use the SR operation on top of that. Since Komondor simulates the actual exchange of frames between nodes in a WLAN, the throughput $S$ experienced by it is directly obtained from:
	\begin{equation}
	S = \frac{[\text{Data bits transmitted successfully}]}{[\text{Total simulation time}]}
	\nonumber
	\end{equation}
	
	The number of data bits (or data packets) transmitted successfully depends on the varying channel conditions and sensed interference. In particular, a given transmission is considered to be successful only if the following conditions hold at the receiver:
	\begin{enumerate}
		\item The power sensed at the receiver from the frame being decoded remains above the CCA/CS.
		\item The Signal-to-Interference-plus-Noise Ratio (SINR) stays above the capture effect (CE) threshold, set to 10 dB. Notice that this is an abstraction of the CE model, which is due to the simplification of the PHY in Komondor.
	\end{enumerate}
	
	\subsection{Scenarios for Evaluation}
\begin{figure}[ht!]
		\centering
		\includegraphics[width=\columnwidth]{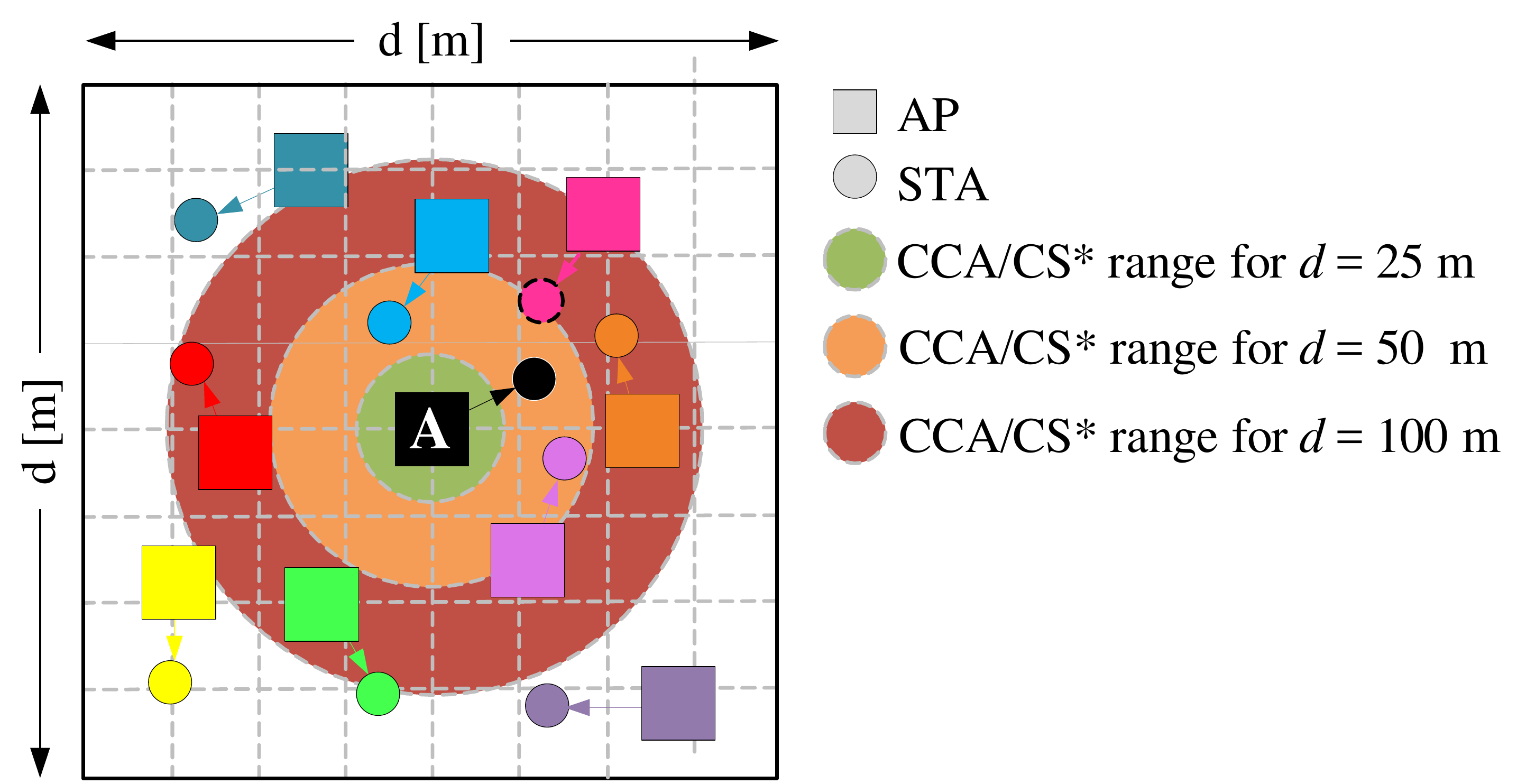}
		\caption{Random deployment with WLAN$_A$ placed in the center. Note that we include the CCA/CS* range of WLAN$_A$ in the map -- corresponding to maximum transmission power (20 dBm) and minimum CCA (-82 dBm) -- for the sake of representing the three considered map densities.}
		\label{fig:random_scenario}
	\end{figure}

	The 11ax SR operation is evaluated in random scenarios like the one depicted in Fig.~\ref{fig:random_scenario}. Notice that, for the sake of illustrating the potential of SR, only the WLAN in the middle (namely, WLAN$_A$) applies the SR operation, while the others remain using the default CCA/CS. WLAN$_A$ is placed at the center of the scenario, so that it is normally exposed to a higher level of interference than the others. We consider that all the WLANs are operating in the same channel, as otherwise, they would not interact.
	
	Different network densities are considered, which are useful to evaluate the gains achieved by the 11ax SR operation in different use cases. In particular, we consider low-density (LD), medium-density (MD), and high-density (HD) scenarios. In summary, we simulate multiple scenarios accounting for $N_m = 3$ maps of sizes $25 \times 25$ , $50 \times 50$, and $100 \times 100$ m$^2$, $N_d = 50$ different random deployments (i.e., node allocation), $N_\text{cca} = 21$ OBSS/PD values ranging from -62 to -82 dBm for WLAN$_A$, and $N_\ell = 16$ traffic loads ranging from $\ell = 1$ to 100 Mbps. In total, $N_s =  N_m \times N_d \times N_\text{cca} \times N_\ell = 50,400$ scenarios (or Komondor inputs) are simulated. The observation time for each simulation is $T = 10$ seconds.
	
	The 11ax PHY and MAC parameters used in the simulations are listed in Table \ref{table:parameters}.
	
	\begin{table}[h]
		\centering
		\caption{Simulation parameters.}
		\resizebox{\columnwidth}{!}{
			\begin{tabular}{c|l|l|}
				\cline{2-3}
				\multicolumn{1}{l|}{}                      & \textbf{Parameter}   & \textbf{Value}
				\\ \hline
				\multicolumn{1}{|c|}{\multirow{9}{*}{PHY}}  
				& Central frequency $f_c$ & 5 GHz \\ 
				\cline{2-3} 
				\multicolumn{1}{|c|}{} & Transmission gain  $G_{tx}$ & 0 dB \\ 
				\cline{2-3}
				\multicolumn{1}{|c|}{} & Reception gain $G_{rx}$ & 0 dB                         \\ 
				\cline{2-3}
				\multicolumn{1}{|c|}{} & Capture Effect threshold $CE$ & 10 dB                         \\ 
				\cline{2-3} 
				\multicolumn{1}{|c|}{} & Path-loss (TMB) $\text{PL}_\text{TMB}(d)$              & See \eqref{eq:tmb}  \\
				\cline{2-3} 
				\multicolumn{1}{|c|}{} & Path-loss intercept $\text{L}_0$              & 54.120 \\
				\cline{2-3} 
				\multicolumn{1}{|c|}{} & Path-loss exponent $\gamma$              & 2.06067  \\
				\cline{2-3} 
				\multicolumn{1}{|c|}{} & Attenuation factor $k$              & 5.25  \\
				\cline{2-3} 
				\multicolumn{1}{|c|}{} & Average num. of walls per m $\overline{W}$              & 0.1467   \\
				\cline{2-3}
				\multicolumn{1}{|c|}{} & Background noise level $N$           & -95 dBm \\
				\cline{2-3}
				\multicolumn{1}{|c|}{} & Legacy OFDM symbol duration $\sigma_\text{leg}$ & 4 \textmu s \\
				\cline{2-3}
				\multicolumn{1}{|c|}{} & OFDM symbol duration (GI-32) $\sigma_{32}$ & 16 \textmu s  \\
				\cline{2-3}
				\multicolumn{1}{|c|}{} & Number of subcarriers (20 MHz)  $N_{sc}$  & 234   \\
				\cline{2-3}
				\multicolumn{1}{|c|}{} & Number of spatial streams $N_{ss}$   & 1  \\
				\cline{2-3}
				\multicolumn{1}{|c|}{} & Transmit power levels $\mathcal{T}$ & 1 to 20 dBm (1 dBm steps) \\
				\hline
				\multicolumn{1}{|c|}{\multirow{16}{*}{MAC}} & Empty slot duration  $\text{T}_e$  & 9 $\mu$s\\ 
				\cline{2-3} 
				\multicolumn{1}{|c|}{} & SIFS duration $T_\text{SIFS}$ & 16 \textmu s  \\
				\cline{2-3} 
				\multicolumn{1}{|c|}{} & DIFS/AIFS duration $T_\text{DIFS/AIFS}$ & 34 \textmu s \\
				\cline{2-3} 
				\multicolumn{1}{|c|}{} & PIFS duration $T_\text{PIFS}$ & 25 \textmu s \\
				\cline{2-3} 
				\multicolumn{1}{|c|}{} & Legacy preamble duration $T_\text{PHY-leg}$ & 20 \textmu s  \\
				\cline{2-3}
				\multicolumn{1}{|c|}{} & HE single-user field duration $T_\text{HE-SU}$ & 100 \textmu s \\
				\cline{2-3} 
				\multicolumn{1}{|c|}{} & ACK duration $T_\text{ACK}$ & 28 \textmu s\\
				\cline{2-3} 
				\multicolumn{1}{|c|}{} & Block ACK duration $T_\text{BACK}$ & 32 \textmu s \\
				\cline{2-3} 
				\multicolumn{1}{|c|}{} & Size OFDM symbol (legacy) $L_{s,l}$ & 24 bits \\
				\cline{2-3} 
				\multicolumn{1}{|c|}{} & Length of data packets $\text{L}_{d}$ & 12,000 bits \\
				\cline{2-3} 
				\multicolumn{1}{|c|}{} & Max. No. of frames in an A-MPDU $N_{\text{agg}}$ & 64 \\
				\cline{2-3} 
				\multicolumn{1}{|c|}{} & Length of an RTS packet $L_\text{RTS}$  & 160 bits \\
				\cline{2-3} 
				\multicolumn{1}{|c|}{} & Length of a CTS packet $L_\text{CTS}$ & 112 bits \\
				\cline{2-3} 
				\multicolumn{1}{|c|}{} & Length of service field $L_\text{SF}$ & 16 bits  \\
				\cline{2-3} 
				\multicolumn{1}{|c|}{} & Length of MAC header $L_\text{MH}$ & 320 bits \\
				\cline{2-3} 
				\multicolumn{1}{|c|}{} & Max. contention window (fixed) $\text{CW}$ & 15 \\
				\cline{2-3} 
				\multicolumn{1}{|c|}{} & Allowed sensitivity levels $\mathcal{S}$ & -82 to -62 (1 dBm steps) \\
				\hline
				\multicolumn{1}{|c|}{\multirow{2}{*}{Other}} & Traffic model $\Lambda$ & Downlink\\
				\cline{2-3} 
				\multicolumn{1}{|c|}{} & Traffic load $\ell$ & 1 to 100 Mbps\\ 
				\cline{2-3} 
				\multicolumn{1}{|c|}{} & Map area $A$ & 25x25, 50x50 and 100x100 $\text{m}^2$\\
				\hline
		\end{tabular}}
		\label{table:parameters}
	\end{table}
	
	\section{Performance Evaluation}
	\label{section:perfromance_evaluation}
	
	\begin{figure}[ht!]
		\centering
		\includegraphics[width=\columnwidth]{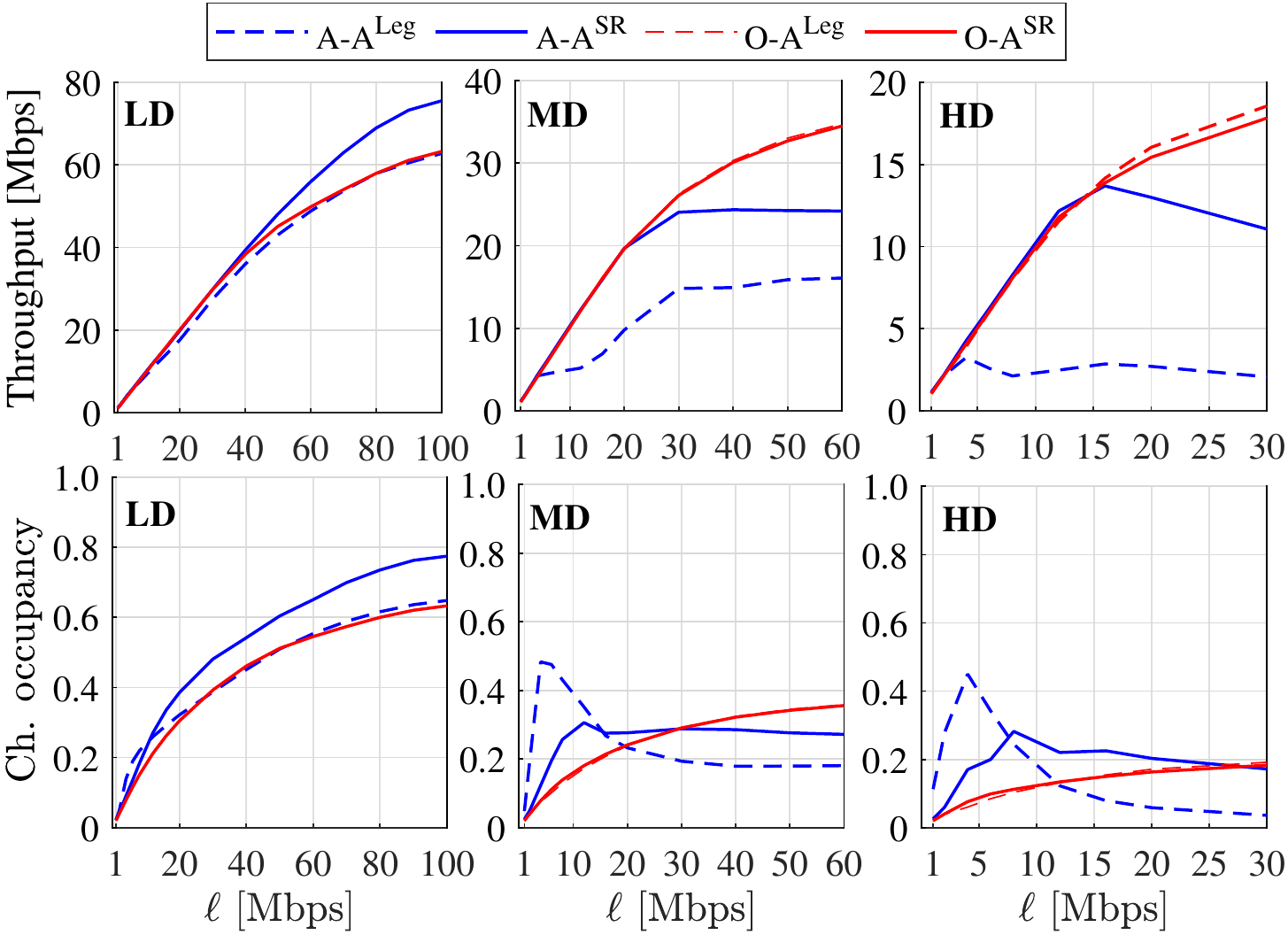}
		\caption{Throughput and channel occupancy experienced by WLAN$_A$ (A) and the other WLANs (O) in low (LD), medium (MD) and high density (HD) deployments. Each curve is named in the legend in the format X-A$^{\text{m}}$, where A$^{\text{m}}$ represents whether WLAN$_A$ uses spatial reuse (SR) or not (Leg).}
		\label{fig:throughput_occupancy}
	\end{figure}
	
	\begin{figure*}[ht!]
		\centering
		\includegraphics[width=0.95\textwidth]{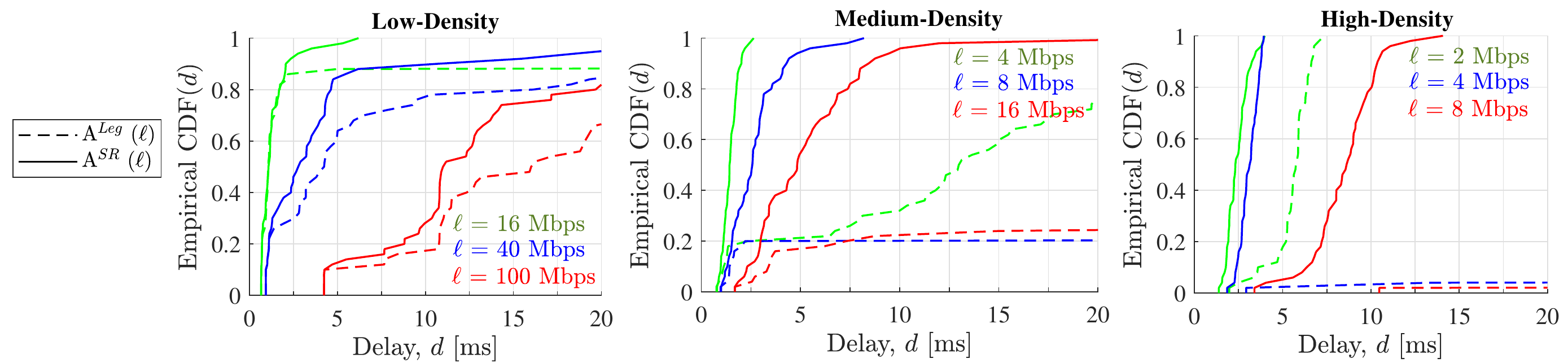}
		\caption{Empirical cumulative distribution function of the average packet delay experienced by WLAN$_A$. Different network densities and traffic loads are considered. Solid and dashed lines indicate whether WLAN$_A$ uses spatial reuse (SR) or not (Leg), respectively.}
		\label{fig:cdf_delay}
	\end{figure*}
	
	Based on the simulation setup described in Section \ref{section:simulation_setup}, we evaluate the potential of the 11ax SR operation in various situations. In particular, for each network density and traffic load, we measure the maximum performance gain that can be achieved by WLAN$_A$ when applying SR. This upper bound is provided by the OBSS/PD value that maximizes the average throughput in each of the 50 random deployments. We also assess the impact of such an optimal SR configuration (from WLAN$_A$'s perspective) on the environment (i.e., the other WLANs). Specifically, we compare the throughput, channel occupancy, and delay obtained by all the other WLANs in two situations: \emph{i)} the legacy CCA/CS is used by the entire network (including WLAN$_A$), and \emph{ii)} Only WLAN$_A$ applies the SR using the optimal OBSS/PD.
	
	Fig.~\ref{fig:throughput_occupancy} shows the highest gains that can be achieved both in throughput and channel occupancy when WLAN$_A$ implements the SR operation. As shown in the first row, significant improvements are achieved on WLAN$_A$'s individual throughput, especially for the highest network density (up to 450\% for medium load). Moreover, importantly, the mean throughput achieved by the rest of WLANs (on average) is barely affected by the SR operation applied by WLAN$_A$.
	
	Regarding channel occupancy (second row), improvements are also noticed as traffic load increases, for every network density. Essentially, more transmission opportunities are achieved due to the lower OBSS/PD threshold used by WLAN$_A$. Nonetheless, an interesting phenomenon occurs for the MD and HD scenarios at low-moderate traffic loads. In these cases, the legacy approach leads to a clear increase on channel occupancy. This increase is caused by the elevated number of re-transmissions performed, which are propitiated by the higher collision probability between overlapping WLANs. We conclude that SR allows using channel resources more efficiently by reducing the contention among neighboring WLANs, and thus boosting parallel transmissions.
	
	Finally, Fig.~\ref{fig:cdf_delay} illustrates the potential reduction on the delay experienced by WLAN$_A$ when implementing SR. In particular, we select the empirical cumulative distribution function (CDF) to highlight the probability of experiencing small and high delays resulting from all the simulated scenarios. Similarly than before, for each scenario, we pick the average delay obtained by the best possible OBSS/PD threshold used by WLAN$_A$ in each of the deployments (in terms of throughput) and compare it with the legacy situation. Three representative traffic load values are included for each network density. As illustrated, the probability of experiencing a high delay rapidly increases with network density and traffic load when the legacy operation is considered. Nevertheless, SR substantially improves delay performance by keeping it at moderate values in most of the simulated scenarios.

	\section{Conclusions}
	In this paper, we introduced the 11ax SR operation and evaluated its potential in a variety of scenarios covering different node densities and traffic loads. To that purpose, we provided an implementation of the SR operation in the 11ax-oriented Komondor simulator. Our results showed that significant gains can be achieved by using the SR operation, especially in high interference situations where both network density and traffic load are high. Therefore, there is a huge potential in SR when it comes to maximizing channel utilization.
	
	As future work, the potential of SR will be studied, especially regarding the interactions that occur when more than one WLAN applies the operation. Moreover, the problem of finding the best OBSS/PD threshold will be studied. In this regard, online learning stands as a powerful and suitable solution, due to the complex inter-WLAN interactions that can be generated by using SR. Promising results have been already shown by applying Machine Learning (ML) to address the SR problem \cite{collaborative, potential}. Finally, the interaction of SR with other techniques included in the 11ax (e.g., directional transmissions, target wake time, OFDMA, etc.) is also worth to be studied.

\end{document}